\documentclass[aps,prc,letterpaper,11pt,twoside,tightenlines,nofootinbib,showpacs]{revtex4}
\usepackage[sort&compress]{natbib}
\usepackage[bookmarks,pdfhighlight=/O,pdfstartview=FitH]{hyperref}
\usepackage{graphicx,mathrsfs,amsmath,amssymb,amsopn,mathbbol,bm}
\usepackage{epstopdf}
\usepackage{color}

\newcommand{\Lag}{\ensuremath{\mathscr{L}}}
\newcommand{\ii}{\ensuremath{\mathrm{i}}}
\makeatletter
  \newcommand{\fslash}[2][0mu]{%
    \mathchoice
      {\fsl@sh\displaystyle{#1}{#2}}%
      {\fsl@sh\textstyle{#1}{#2}}%
      {\fsl@sh\scriptstyle{#1}{#2}}%
      {\fsl@sh\scriptscriptstyle{#1}{#2}}}
  \newcommand{\fsl@sh}[3]{%
    \m@th\ooalign{$\hfil#1\mkern#2/\hfil$\crcr$#1#3$}}
\makeatother

\newcommand{\feynint}[1]{\ensuremath{\int \frac{\mathrm{d}^{d} {#1}}{(2
\pi)^{d}}}}
\DeclareMathOperator{\Tr}{Tr}
\DeclareMathOperator{\re}{Re}
\DeclareMathOperator{\im}{Im}
\renewcommand{\d}{\ensuremath{\mathrm{d}}}
\newcommand{\erw}[1]{\ensuremath {\left \langle {#1} \right \rangle}}
\preprint{nucl-th/0412015}
\begin{document}

\title{Thermalization of Heavy Quarks in the Quark-Gluon Plasma} 
\author{Hendrik van Hees}
\affiliation{Cyclotron Institute, Texas A{\&}M University, College 
Station, Texas 77843-3366, USA}
\author{Ralf Rapp}
\affiliation{Cyclotron Institute, Texas A{\&}M University, 
College Station, Texas 77843-3366, USA}

\date{January 26, 2005}

\begin{abstract}
  Charm- and bottom-quark rescattering in a Quark-Gluon Plasma (QGP) is
  investigated with the objective of assessing the approach toward
  thermalization.  Employing a Fokker-Planck equation to approximate the
  collision integral of the Boltzmann equation we augment earlier studies
  based on perturbative parton cross sections by introducing resonant
  heavy-light quark interactions.  The latter are motivated by recent QCD
  lattice calculations which indicate the presence of ``hadronic'' states
  in the QGP.  We model these states by colorless (pseudo-) scalar and
  (axial-) vector $D$- and $B$-mesons within a heavy-quark effective theory
  framework. We find that the presence of these resonances at moderate QGP
  temperatures substantially accelerates the kinetic equilibration of
  $c$-quarks as compared to using perturbative interactions. We also
  comment on consequences for $D$-meson observables in ultra-relativistic
  heavy-ion collisions.
\end{abstract}

\maketitle

\section{Introduction}

Hadrons containing heavy quarks are valuable probes of the strongly
interacting matter produced in high-energy collisions of heavy nuclei.  The
spectral properties of (bound) $c$-$\bar c$ states, such as their binding
energy and (decay) width, are expected to undergo substantial modifications
in a Quark-Gluon Plasma (QGP) and thus affect charmonium yields and
momentum spectra in heavy-ion reactions with QGP formation (see,
\textit{e.g.}, Refs.~\cite{Vogt99,Satz00,rapgra03} for overviews).
Reinteractions of {\em individual} $c$-quarks in the QGP will reflect
themselves in transverse-momentum ($p_T$-) spectra of open charm hadrons
($D$-mesons)~\cite{kar03,djo04,ASW04,bra04}, most notably their elliptic
flow, $v_2(p_T)$, in semi-central collisions~\cite{Bats03,GKR04}.
Preliminary experimental results from the Relativistic Heavy-Ion Collider
(RHIC) indicate the possibility that the $D$-meson $v_2$ could be similar
in magnitude to the one of light hadrons~\cite{phenix-e04,star-e04}. Since
the $c$-quark is rather heavy, this would be quite remarkable and could
provide important insight into (nonperturbative) properties of the QGP at
moderate temperatures, $T\simeq 1$-$2 \; T_c$. \textit{E.g.}, in the
light-quark sector, parton rescattering through hadron-like states in the
QGP (motivated by recent lattice calculations of Quantum Chromodynamics
(QCD) at finite temperature) has been suggested as a mechanism to enhance
partonic cross sections~\cite{SZ04,BLRS04,Shakin03} in order to facilitate
rapid thermalization of the bulk matter at RHIC as required in
hydrodynamical models.  The notion of charmonium resonances in the
QGP~\cite{Datta02,Umeda02} has been applied earlier to assess $J/\psi$
production at SPS and RHIC~\cite{rapgra04b}. Employing kinetic rate
equations to consistently account for both dissociation and regeneration
reactions, $c+\bar c \leftrightarrow J/\psi+X$, $J/\psi$ resonance
formation in the QGP via $c$-$\bar c$ ``coalescence'' turns out to be the
dominant contribution to the final yield in central Au-Au collisions at
RHIC~\cite{Thews01,Ko02,rapgra04b} (see also Refs.~\cite{pbm00,Goren01}).
Furthermore, $J/\psi$ production was found to be sensitive to the in-medium
properties of charm quarks, \textit{i.e.}, their in-medium masses and
degree of kinetic equilibration~\cite{Thews03,rapgra02,rapgra04b}.
 
Early studies of charm-quark thermalization in the QGP have been conducted
in Ref.~\cite{Svet88} based on elastic perturbative QCD (pQCD) cross
sections~\cite{com79}, $c + q (\bar q, g) \to c + q (\bar q, g)$,
implemented into a Fokker-Planck equation to approximate the collision
integral of the Boltzmann equation (see Ref.~\cite{MT03} for a recent
application to light partons). With a strong coupling constant
$\alpha_s$=$0.6$ rather short relaxation times of around $\sim$$4$~fm/$c$
have been found for a massless QGP at temperatures $T$$\simeq$400~MeV.
However, since the relaxation times are essentially proportional to
$\alpha_s^{-2}$, more moderate values of $\alpha_s$ (\textit{e.g.}, $0.3$)
lead to a significant increase (factor 3-4), rendering thermalization of
$c$-quarks under RHIC conditions unlikely~\cite{MS97}.  This is also
consistent with recent transport studies~\cite{Chen04,Mol04}.

In the present article we evaluate heavy-quark rescattering in the QGP 
via ``$D$"- and ``$B$"-meson resonances, the existence of which is the 
main assumption in our work. Although lattice QCD has not yet addressed
in-medium heavy-light ($Q$-$\bar q$) spectral functions at finite 
temperature, resonance-like correlations in the QGP are quite plausible 
in view of the indications for both $q$-$\bar q$~\cite{AH03,KL03} and 
$Q$-$\bar Q$ systems. Further support for this assumption  
is provided by calculations using effective four-quark
interactions within Nambu-Jona-Lasinio
models~\cite{GK92,Blasch02,Blasch03}. With form factors and coupling
constants adjusted to free $D$-meson masses, finite-temperature
calculations lead to resonances above the $Q$-$\bar q$ threshold 
in the QGP, with appreciable widths of several hundred MeV.

Our article is organized as follows: in Sec.~\ref{sec_Dmes} we introduce
Lagrangians for resonant $c$-$q$ interactions based on chiral and
heavy-quark symmetry for pseudo-/scalar and axial-/vector multiplets
(Sec.~\ref{sec_lag}) and evaluate pertinent scattering amplitudes
(Sec.~\ref{sec_ampl}). In Sec.~\ref{sec_rescat} we apply the resulting
cross sections within a Fokker-Planck equation; we first determine
temperature and momentum dependencies of drag and diffusion constants in a
static QGP (Sec.~\ref{sec_fp}), and then evaluate the time evolution of
$c$-quark transverse-momentum ($p_T$) spectra within an expanding fireball
model to simulate conditions in heavy-ion collisions at RHIC
(Sec.~\ref{sec_evo}). We conclude in Sec.~\ref{sec_concl} including a
discussion on open-charm observables in ultrarelativistic heavy-ion
collisions (URHICs).

%%%%%%%%%%%%%%%%%%%%%%%%%%%%%%%%%%%%%%%%%%%%%%%%%%%%%%%%%%%%%%%%%%
\section{$D$-Mesons in the Quark-Gluon Plasma}
\label{sec_Dmes}
%%%%%%%%%%%%%%%%%%%%%%%%%%%%%%%%%%%%%%%%%%%%%%%%%%%%%%%%%%%%%%%%%%

%%%%%%%%%%%%%%%%%%%%%%%%%%%%%%%%%%%%%%%%%%%%%%%%%%%%%%%%%%%%%%%%
\subsection{Heavy-Light Quark Lagrangians}
\label{sec_lag}
%%%%%%%%%%%%%%%%%%%%%%%%%%%%%%%%%%%%%%%%%%%%%%%%%%%%%%%%%%%%%%%%

Our description of $D$-meson resonances in the QGP is based on a rather
simplistic quark-meson model, accounting, however, for the relevant
symmetries, \textit{i.e.}, chiral symmetry in the light quark sector
($u$-$d$) and heavy-quark symmetry for $c$-quarks. As a minimal
set of resonances, consistent with lattice calculations~\cite{AH03}, we
assume the lowest-lying pseudoscalar ($D$) and vector mesons ($D^*$) to
survive above $T_c$\footnote{In the vacuum these are associated with
  $D^+$(1870), $D^0(1865)$ and $D^*$(2010) mesons.}. In addition,
(approximate) restoration of chiral symmetry mandates the existence of
pertinent $SU(2)_f$ chiral partners in the scalar ($D_0^*$) and
axial-vector ($D_1$) channel with mass and width identical to $D$ and
$D^*$, respectively. The effective Lagrangian thus takes the form
\begin{equation}
\begin{split}
\label{lag-d}
\Lag_{Dcq} =& \Lag_D^0 + \Lag_{c,q}^0 - \ii G_S \left( \bar q \Phi_0^*
\frac{1+\fslash{v}}{2} c - \bar q \gamma^5 \Phi \frac{1+\fslash{v}}{2}
 c + h.c. \right)
\\
& - G_V \left( \bar q \gamma^{\mu} \Phi_{\mu}^* \frac{1+\fslash{v}}{2} c -
  \bar q \gamma^5 \gamma^{\mu} \Phi_{1\mu} \frac{1+\fslash{v}}{2} c + h.c.
\right)
\end{split}
\end{equation}
with the usual free (kinetic and mass) terms for quarks  
and $D$-mesons,
\begin{equation}
\begin{split}
\Lag_{c,q}^0 &= \bar{c}(\ii \fslash{\partial}-m_c) c+\bar{q}\ii
\fslash{\partial} q,\\
\Lag_D^0 & =  (\partial_{\mu} \Phi^{\dagger})(\partial^{\mu} \Phi) +
(\partial_{\mu} {\Phi_0}^{*\dagger})(\partial^{\mu} \Phi_0^*)
-m_S^2(\Phi^{\dagger} \Phi+\Phi_0^{*\dagger} \Phi_0^*) \\
& \quad -\frac{1}{2} (\Phi_{\mu \nu}^{*\dagger} \Phi^{*\mu \nu} 
+ \Phi_{1 \mu \nu}^{\dagger}
\Phi_1^{\mu \nu}) + m_V^2 (\Phi_{\mu}^{*\dagger} \Phi^{*\mu} + 
\Phi_{1 \mu}^{\dagger} \Phi_1^{\mu}) \ . 
\end{split} 
\end{equation}
The fields $\Phi$ represent {\em anti}-$D$-mesons, transforming as
isospinors under isospin rotations.

The interaction terms in Eq.~(\ref{lag-d}) will be evaluated 
to leading order in $1/m_c$ according to heavy-quark effective 
theory (HQET)~\cite{ebert94}. This ensures the absence of 
unphysical (4-D longitudinal) degrees of freedom for massive (axial-)
vector meson fields as encoded in the transversality constraints
\begin{equation}
v_{\mu} \Phi^{*\mu} = v_{\mu} \Phi_1^{\mu}=0 \ 
\label{trans}
\end{equation}
($v_\mu$: four velocity of the charm quark or $D$-meson). 
Equivalent relations hold for
the $D$-meson self-energies (to leading order in the $1/m_c$ expansion of
HQET).

In the light-flavor sector the relevant symmetry is the invariance of 
the Lagrangian under the chiral $SU(2)_L\times SU(2)_R$ group. For
infinitesimal angles $\delta\vec\phi_{V,A}$, it is characterized by the
standard vector and axial-vector transformations acting on the quark 
fields as
\begin{equation}
\label{chiop}
q \rightarrow (1+\ii \delta \vec{\phi}_V \vec{t} + \ii \delta \vec{\phi}_A
\vec{t} \gamma_5) q, \quad c \rightarrow c 
\end{equation}
with $\vec{t}=\vec{\tau}/2$ the generators of $SU(2)$ ($\vec{\tau}$:
Pauli matrices). To construct the group operations for the $\bar D$  
fields, we identify the latter with underlying quark currents
according to
\begin{equation}
\label{meson-trafos}
\Phi \sim \bar{c} \gamma_5 q \ , \quad \Phi_0^* \sim \bar{c} q \ ,
\quad \Phi^*_\mu \sim  \bar{c} \gamma_\mu q \ , 
\quad \Phi_{1,\mu} \sim \bar{c} \gamma_\mu \gamma_5 q \ . 
\end{equation}
Here, ``$\sim$" denotes ``transforms under (\ref{chiop}) like''. Thus the
transformation rules for the $\bar{D}$-meson fields follow as
\begin{equation}
\begin{split}
\label{meson-trafos2}
\Phi &\rightarrow \Phi+\ii \delta \vec{\phi}_A \cdot \vec{t} \Phi + \ii \delta
\vec{\phi}_V \cdot \vec{t} \Phi_0^*, \\
\Phi_0^* &\rightarrow \Phi_0^*+\ii \delta \vec{\phi}_A \cdot \vec{t}
\Phi_0^* + \ii \delta \vec{\phi}_V \cdot \vec{t} \Phi \ .
\end{split}
\end{equation}
With these properties the Lagrangian, Eq.~(\ref{lag-d}), is a scalar under
chiral transformations.

For $D_s$-mesons we restrict ourselves to the (experimentally known) 
pseudoscalar and vector states since spontaneous chiral symmetry 
breaking in the strange-quark sector is expected to persist to 
temperatures significantly larger than $T_c$ (characterized by sizable 
values of the strange condensate $\langle\bar ss\rangle$). The 
corresponding Lagrangian is therefore taken to be
\begin{equation}
\label{lag-ds}
\begin{split}
  \Lag_s = &(\partial_{\mu} \Phi_s^{\dagger}) (\partial^{\mu} \Phi_s)-m_{D_s}^2
  \Phi_s^{\dagger} \Phi_s -\frac{1}{2} \Phi_{s\mu\nu}^{*\dagger} \Phi_s^{\mu\nu} 
  +m_{D_s^*}^2 \Phi_{s\mu}^{*\dagger} \Phi_{s}^{\mu} \\
 & -\ii G_{s,S} \bar{s} \gamma^5 \Phi_s \frac{1+\fslash{v}}{2} c - 
        G_{s,V} \bar{s} \gamma^{\mu}  
 \Phi_{s,\mu}^* \frac{1+\fslash{v}}{2} c \ .
\end{split}
\end{equation}

Eqs.~(\ref{lag-d}) and (\ref{lag-ds}) constitute our basic vertices for
effective charm-light quark interactions in the QGP (analogous expressions
hold in the $b$-quark sector upon the replacement $c\to b$ and $D\to B$).
The underlying parameters, \textit{i.e.}, the bare masses of the $D$-meson
and their coupling strength to the quarks, will be fixed to resemble more
microscopic model calculations of corresponding spectral functions above
$T_c$, as discussed below.  In the following we will also impose the spin
symmetry of HQET implying that the spectral properties of pseudoscalar and
vector mesons, and consequently their masses and couplings to quarks, are
equal.

%%%%%%%%%%%%%%%%%%%%%%%%%%%%%%%%%%%%%%%%%%%%%%%%%%%%%%%%%%%%%%%%%%%%%%%%%
\subsection{Meson Self-Energies and Heavy-Light Quark Scattering 
Amplitudes}
\label{sec_ampl}
%%%%%%%%%%%%%%%%%%%%%%%%%%%%%%%%%%%%%%%%%%%%%%%%%%%%%%%%%%%%%%%%%%%%%%%%%

\begin{figure}
\begin{minipage}{7.6cm}
\includegraphics[width=7.5cm]{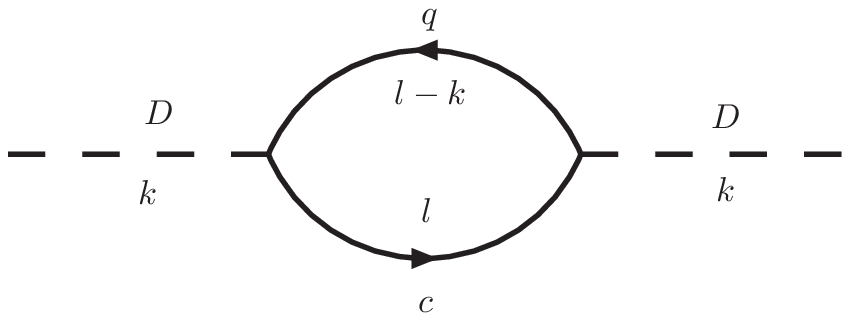}
\end{minipage}
\begin{minipage}{5.1cm}
\includegraphics[width=5cm]{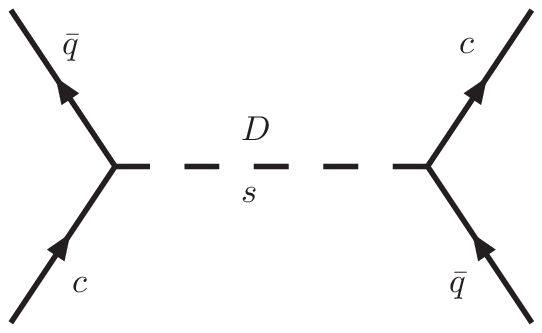}
\end{minipage}
\begin{minipage}{3.1cm}
\includegraphics[width=3cm]{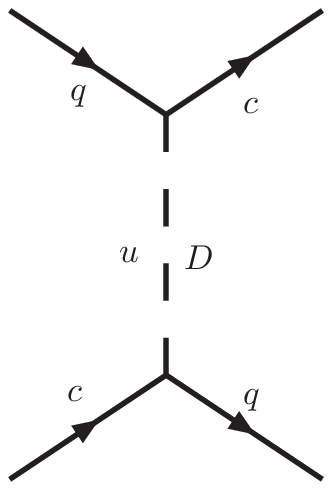}
\end{minipage}
\caption{Left panel: $c$($b$)-$q$ loop diagram representing the 
$D$($B$)-meson self-energy in the QGP. 
Right panel: "meson"-exchange diagrams 
contributing  to the invariant matrix elements for the scattering of 
charm quarks on light quarks ($u$-channel) and anti-quarks ($s$-channel).}
\label{fig_diag} 
\end{figure}

The key ingredient for the heavy-quark scattering amplitudes are
the heavy-meson exchange propagators,
\begin{equation}
\label{prop}
D_{D,B}(k)=\frac{1}{k^2-m_{D,B}^2-\Pi_{D,B}(k)} \ , 
\end{equation}
which are essentially determined by the underlying self-energies,
$\Pi_{D,B}$, together with the bare resonance masses, $m_{D,B}$, of the
Lagrangian. In the following, we will evaluate the self-energies in 
terms of the  heavy-light-quark loop (cf.~the left diagram in 
Fig.~\ref{fig_diag}), thereby investigating two different schemes to
regularize the divergent loop integrals to assess the robustness of our
results. We will concentrate again on the charm-quark case, but 
completely analogous expressions apply to the bottom sector.

Within the dimensional regularization scheme (see, \textit{e.g.},
\cite{ram89}), the interaction vertices of Eq.~(\ref{lag-d}) yield a
self-energy for (pseudo-) scalar $D$-mesons of four-momentum $k$ of the form
\begin{align}
\Pi_D(s) = \Pi_{D_0^*}(s) &= 
3 \ii G^2 \mu^{4-d} \feynint{l} \Tr [\gamma_5 G_q(l+k) \gamma_5 G_Q(l)] 
\nonumber\\
&= \frac{3 G^2}{8\pi^2} \Bigg \{
   \frac{4 m_c^2-2 s}{4-d} + 3 m_c^2-2s + (s-2 m_c^2) \left [\gamma + \ln
     \left (\frac{m_c^2-s}{4 \pi \mu^2} \right ) \right ] 
\nonumber\\ 
  &  \quad \qquad + \frac{m_c^4}{s}
   \ln \left (\frac{m_c^2-s}{m_c^2} \right ) \Bigg \} \ , 
\label{self}
\end{align}
($s=k^2$). The first term contains the quadratic divergence for $d
\rightarrow 4$, showing that the self-energy can be rendered finite with a
field- and mass-renormalization; $\mu$ denotes the mass scale in the
dimensional-regularization scheme to keep the momentum dimensions of the
integrals as for $d=4$, and $\gamma \simeq 0.577$ is Euler's constant. The
(regularization-independent) imaginary part is given by
\begin{equation}
\label{imse}
\im \Pi_D(s)=-\frac{3G_S^2}{8\pi} \frac{(s-m_c^2)^2}{s} 
\Theta(s-m_c^2) \ .
\end{equation}
For the vector and axial-vector mesons, we employ the HQET propagator for
the charm quark, 
\begin{equation}
G_v(l)=\frac{m_c}{m_c v \cdot l + \ii \eta} \frac{1+\fslash{v}}{2} \ ,  
\label{Gv}
\end{equation}
to obtain a transverse self-energy in leading order of the expansion in
$1/m_c$.  In Eq. (\ref{Gv}), the ``residual momentum'' $l$ of the charm
quark is defined via its total four-momentum as $l_c=m_c v+l$. For the
$D$-meson fields the corresponding decomposition for the heavy-quark
expansion reads
\begin{equation}
k_c=m_c v+k, \quad s=k_c^2=m_c^2 \left [1 + \frac{2 v \cdot k}{m_c} + 
    \mathcal{O}\left (\frac{p^2}{m_c^2} \right)  \right ] \; 
    \Rightarrow \; v \cdot k = \frac{s-m_c^2}{2 m_c} + \mathcal{O}\left
    (\frac{p^2}{m_c^2} \right) \ .
\end{equation}
Applying the Feynman rules, and after some (Dirac) algebra, the 
dimensionally regularized polarization tensor for the axial-/vector
becomes
\begin{equation}
\Pi_{D^*}^{\mu \nu}(p)=\Pi_{D_1}^{\mu \nu}(p)= -6 \ii G_V^2 \mu^{4-d} m_c
\int \frac{\d^{d} l}{(2 \pi)^d} \frac{(l+p)_{\nu} v_{\mu} - 
(l+p) \cdot v g_{\mu \nu}}{[(l+p)^2+\ii \eta](m_c v \cdot l + \ii \eta)} \ .
\end{equation}
The integral is conveniently evaluated with help of the 
identity~\cite{georgi91} 
\begin{equation}
\frac{1}{a b}=\int_{0}^{\infty} \d \lambda \frac{2}{(a+2 \lambda b)^2} \ .
\end{equation}
Upon integrating over $\lambda$ one obtains for the imaginary part 
\begin{equation}
\im \Pi_{D^*}^{\mu \nu}(p)= \im \Pi_{D_1}^{\mu \nu}(p) 
= -(v_{\mu} v_{\nu}-g_{\mu
    \nu}) \frac{3G_V^2}{8 \pi} \frac{(s-m_c^2)^2}{m_c^2} 
  \Theta(s-m_c^2) \ . 
\end{equation}
Since up to corrections ${\cal O}(v k/m_c)$ one can identify $m_c^2=s$ 
in the denominator, one finds 
\begin{equation}
\label{vecse}
\Pi_{D^* \mu \nu}(s) = \Pi_{D_1 \mu \nu}(s)=(v_{\mu} v_{\nu} - g_{\mu \nu})
\Pi_{D}(s) \ , 
\end{equation}
which is the expected result from spin symmetry of the HQET.
We define the renormalization constants for the self-energy by the 
following conditions, 
\begin{equation}
\label{rencond}
\partial_s \Pi_{D}^{(\text{ren})}(s)|_{s=0}=0 \ , \quad 
\re \Pi_{D}^{(\text{ren})}(s)|_{s=m_D^2}=0 \ . 
\end{equation}
The first condition ensures that, within a vector dominance model, the
photon propagator has residuum of unity at $s=0$, while the second one
implies that the renormalized meson mass coincides with the bare mass of
the Lagrangian. The renormalized self-energy is, of course, independent
of the dimensional-regularization scale $\mu$.

As an alternative way of regularizing the divergent self-energy 
integrals we introduce a dipole form factor at the $c$-$q$-$D$ vertex,
\begin{equation}
\label{dipolff}
F(|\vec{q}|)=
\left ( \frac{2 \Lambda^2}{2 \Lambda^2+\vec{q}^2} \right )^2 \ ,
\end{equation}
where $\vec{q}$ denotes the three-momentum of the quarks in the 
center-of-mass frame. The imaginary part of the self-energy
is then given by
\begin{equation}
\label{dipse}
\im \Pi_{D}^{(\text{ff})}(s)= \im \Pi_{D}(s) F^2(|\vec{q}|) \ ,
\end{equation}
$|\vec{q}|=(s-m_c^2)/(2\sqrt{s})$, while the real part is determined 
by an unsubtracted dispersion integral.
The bare meson mass is then adjusted to render a vanishing real part of 
the propagator at the physical resonance mass.

As default parameters for our calculations we use massless light quarks, 
a charm-quark mass of $m_c=1.5$~GeV and physical $D$-meson masses
$(m_D^{\text{phys}})^2=m_D^2-\re \Pi_D[(m_D^{\text{phys}})^2]=2$~GeV
(corresponding to a vanishing real part in the propagator). The coupling
constant $G\equiv G_{S,V}$ is varied to allow for widths of the $D$-meson
spectral functions of $300$-$500$~MeV, to approximately cover the
range suggested by effective quark models~\cite{GK92,Blasch02,Blasch03}.
It is important to note that we assume the $D$-meson resonances to be
located \emph{above} the $c$-$\bar q$ mass threshold, $m_c$+$m_{\bar q}$, 
which renders them accessible in $c$-$\bar q$ scattering processes. 
The situation is quite different for (bound) meson states 
(\textit{i.e.}, below the
anti-/quark threshold), where the resonant part of the scattering amplitude
cannot be probed through $c+\bar q\to c+\bar q$ interactions (even for
resonance masses close to threshold, thermal energies of anti-/quarks imply
that the average collision energy is significantly above the resonance
peak). In this case, other processes need to be calculated, \textit{e.g.},
$c+\bar q \to D+g$, where the extra gluon in the final state carries away
four-momentum to allow the $D$-meson to emerge on-shell.  The same framework
is also applied to the bottom sector, with $b$-quark and $B$-meson masses
of $m_b=4.5$~GeV and $m_B=5$~GeV, respectively.

As we will see below, for equal on-shell masses and widths of the $D$-meson
spectral functions, both regularization schemes lead to quite comparable
results for the thermal relaxation properties of $c$-quarks, even though
the off-shell properties differ significantly.

The $D$-meson propagators figure into the
invariant matrix elements for elastic $c$-quark scattering off quarks, 
$c+q \rightarrow c+q$, and antiquarks, $c+\bar{q} \rightarrow c+\bar{q}$,
in the $u$- and $s$-channel, respectively. One
finds
\begin{align}
\label{mel}
\sum |\mathcal{M}_{\bar{q}}|^2 &= 720 G^4 (s-m_c^2)^2
\left |D_{D}(s) \right|^2
\\
\sum |\mathcal{M}_{q}|^2 &= 720 G^4 (u-m_c^2)^2 \left |D_{D}(u)
\right|^2 \ ,  
\end{align}
where we have summed over the contributions of all light-quark 
resonances, being equal due to either the heavy-quark symmetry 
($D$-$D^*$, $D_0^*$-$D_1$) or chiral symmetry ($D$-$D_0^*$, 
$D^*$-$D_1$). We also included finite-mass strange quarks along with 
$D_s$- and $D_s^*$-mesons.

In addition to resonant interactions, elastic scattering in 
pQCD is accounted for to leading order in $\alpha_s$, 
${\cal O}(\alpha_s^2)$. The corresponding matrix elements~\cite{com79} 
have been supplemented with an additional gluon Debye (screening) mass, 
$\mu_g=g T$,
which regulates the forward singularity in the $t$-channel exchange
graphs~\cite{Svet88}. The strong coupling constant will be varied over
the range $\alpha_s=$0.3-0.5.

%%%%%%%%%%%%%%%%%%%%%%%%%%%%%%%%%%%%%%%%%%%%%%%%%%%%%
\section{Charm Quark Rescattering in the QGP}
\label{sec_rescat}
%%%%%%%%%%%%%%%%%%%%%%%%%%%%%%%%%%%%%%%%%%%%%%%%%%%%%%

%%%%%%%%%%%%%%%%%%%%%%%%%%%%%%%%%%%%%%%%%%%%%%%%%%%%%%%%%%%%%%%%%%%%%
\subsection{Fokker-Planck Equation, Drag and Diffusion Coefficients}
\label{sec_fp}
%%%%%%%%%%%%%%%%%%%%%%%%%%%%%%%%%%%%%%%%%%%%%%%%%%%%%%%%%%%%%%%%%%%%%

The above matrix elements are now implemented within a kinetic theory
framework to assess the thermalization time scales for heavy quarks in a
QGP. Following the steps outlined in Ref.~\cite{Svet88}, we start from the
Boltzmann equation for the heavy-quark distribution function $f(t,\vec p)$ and neglect any mean-field terms.  Furthermore assuming the
scattering processes to be dominated by small momentum transfers one
arrives at a Fokker-Planck equation describing the time evolution of $f$ in
momentum space,
\begin{equation}
\label{fp-eq}
\frac{\partial f(t,\vec{p})}{\partial t} = \frac{\partial}{\partial p_i}
\left [A_i(\vec{p}) + \frac{\partial}{\partial p_j} B_{ij}(\vec{p}) \right
] f(t,\vec{p}) \ .
\end{equation}
For an isotropic (rotationally invariant) plasma the drag and 
diffusion coefficients in (\ref{fp-eq}) can be decomposed as  
\begin{align}
A_i(\vec{p})&=p_i A(|\vec{p}|)
\\
B_{ij}(\vec{p}) &= \left ( \delta_{ij}
 - \frac{p_i p_j}{\vec{p}^2} \right) B_0(|\vec{p}|) + \frac{p_i
 p_j}{\vec{p}^2} B_1(|\vec{p}|)
\end{align}
with the scalar functions 
\begin{align}
A(|\vec{p}|) &= \erw{1} -\frac{\erw{\vec{p} \cdot \vec{p}{\,}'}}{\vec{p}^2}
\label{A}
\\
B_0(|\vec{p}|) &= \frac{1}{4} \left [
  \erw{\vec{p}{\,}'{}^2}-\frac{\erw{(\vec{p} \cdot \vec{p}{\,}')^2}}{\vec{p}^2}
\right]
\label{B0}
\\
B_1(|\vec{p}|) &= \frac{1}{2} \left[
\frac{\erw{(\vec{p} \cdot \vec{p}{\,}')^2}}{\vec{p}^2} 
-2 \erw{\vec{p}{\,}' \cdot \vec{p}} + \vec{p}^2 \erw{1} \right] \ . 
\label{B1}
\end{align}
The averaging is defined by
\begin{align}
  \erw{X(\vec{p}{\,}')} &= \frac{1}{2 E_p} \int \frac{\d^3 \vec{q}}{(2 \pi)^3
    2E_q} \int \frac{\d^3 \vec{q}{\,}'}{(2 \pi)^3 2E_{q'}} \int \frac{\d^3
    \vec{p}{\,}'}{(2\pi)^3 2E_{p'}} \frac{1}{\gamma_c} \sum |\mathcal{M}|^2
  \nonumber\\
  & \quad \times (2 \pi)^4 \delta^{(4)}(p+q-p'-q') \hat{f}(\vec{q})
  X(\vec{p}{\,}') \ ,
\label{erw}
\end{align}
where $\vec{p}$ ($\vec{p}{\,}'$) and $\vec{q}$ ($\vec{q}{\,}'$) denote the momenta
of the incoming (outgoing) charm- and light-quark/gluon, respectively, and
$\hat{f}(\vec{q})$ are the thermal Maxwell-Boltzmann distribution functions
of the light partons\footnote{As pointed out in Ref.~\cite{MS97} the use of
  quantum distribution functions (Bose-Einstein and Fermi-Dirac) induces
  moderate corrections to the drag and diffusion coefficients obtained
  from pQCD rescattering; this is still true for the ``D"-meson resonance
  interactions as they involve contributions from thermal anti-/quarks
  only which, at the temperatures under consideration, are mostly
  non-degenerate. {\it E.g.}, at $T$=300~MeV, the deviations when using
  Fermi distributions amount to $\sim$20\%, which is inside the
  uncertainties associated with the resonance parameters as discussed
  below. For the total coefficients (including both pQCD and resonance 
  rescattering) the effects of quantum statistics approximately cancel.
  In the following, we will use Maxwell-Boltzmann distributions for light 
  partons, since  (i) it is more consistent within the Fokker-Planck
  treatment for the heavy-quark, the thermal limit of which is a Boltzmann
  distribution, and    
  (ii) it minimizes the deviations in the dissipation-fluctuation theorem, 
   cf. the discussion following Eq.~(\ref{diss-fluct}) below.}.
After integrating over the four-momentum conserving $\delta$-function, 
and making use of Lorentz invariance of the matrix elements, the 
expressions for the scalar coefficients, Eqs.~(\ref{A})-(\ref{B1}), 
can be reduced to numerically tractable three-dimensional 
integrals\footnote{We note that, when using the pQCD matrix elements, 
  our final expressions are larger by a factor of 2
  than the ones originally derived in Ref.~\cite{Svet88}, which confirms
  the findings of Ref.~\cite{MS97}.}.
\begin{figure}[!t]
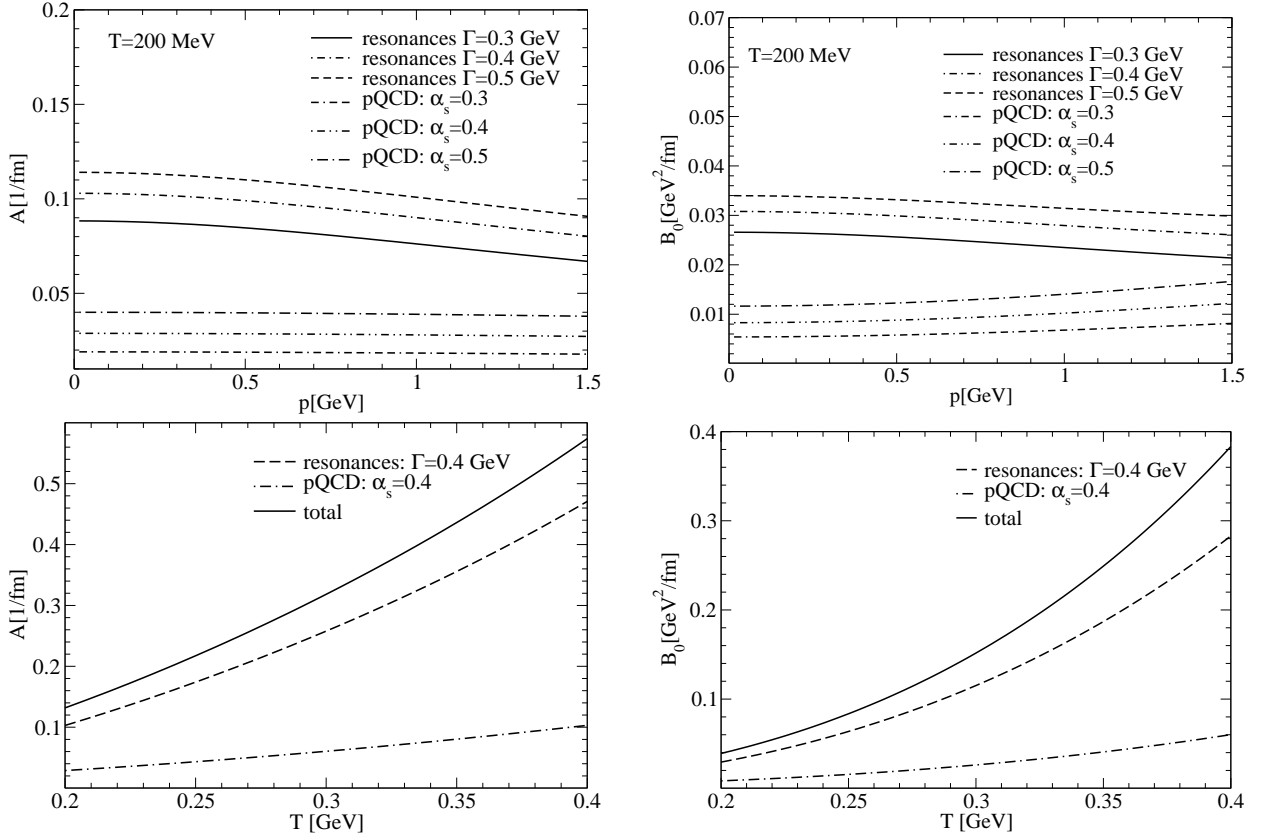

\begin{minipage}{0.48\textwidth}
\includegraphics[width=\textwidth]{reso-model-A}
\end{minipage}\hfill
\begin{minipage}{0.48\textwidth}
\includegraphics[width=\textwidth]{reso-model-B0}
\end{minipage}\vspace*{0.5mm}
\begin{minipage}{0.48\textwidth}
\includegraphics[width=\textwidth]{reso-model-A-vsT}
\end{minipage}\hfill
\begin{minipage}{0.48\textwidth}
\includegraphics[width=\textwidth]{reso-model-B0-vsT}
\end{minipage}
\caption{Upper panel: drag coefficient $A$ (left) and diffusion coefficient
  $B_0$ (right) as a function of $c$-quark three-momentum at a temperature
  of $T=200$~MeV for various values of $\alpha_s$ (pQCD scattering) and
  $D$-meson widths (resonance exchanges).  Lower panel: The same quantities
  as a function of temperature at fixed three-momentum $|\vec p|=0$.}
\label{fig_coeff}
\end{figure}
The temperature and momentum dependencies of the
$A$ and $B_0$ coefficients are summarized in Fig.~\ref{fig_coeff}. The main
finding here is that for resonant rescattering both are increased by a
substantial factor ($\sim$3) over the pQCD results (the latter are very
similar to Ref.~\cite{Svet88} as the extra factor of 2 is essentially
compensated by the larger screening mass, $\mu_g=gT$, used in our
calculation). The main reason for this effect is not so much an increase in
the total cross section (the peak cross section in the resonance case of
about $10$~mb is comparable to an almost constant $4$~mb for pQCD), but the
isotropic angular distribution of the resonance cross sections in contrast
to forward dominated pQCD rescattering.  Also note that the coefficients
are rather insensitive to the underlying coupling constants for both
resonance and pQCD scattering; naively, the matrix elements are
proportional to the 4th power of the coupling constant, which would imply a
variation by a factor of $(5/3)^2$$\simeq$2.6 for the coefficients $A$ and
$B_0$ for the parameter ranges shown in Fig.~\ref{fig_coeff}. The much
smaller actual variation is due to compensating effects induced by an
increased Debye mass for pQCD $t$-channel gluon exchange, and by an
increased resonance width for $s$-channel $D$-meson exchange.
\begin{figure}[!t]
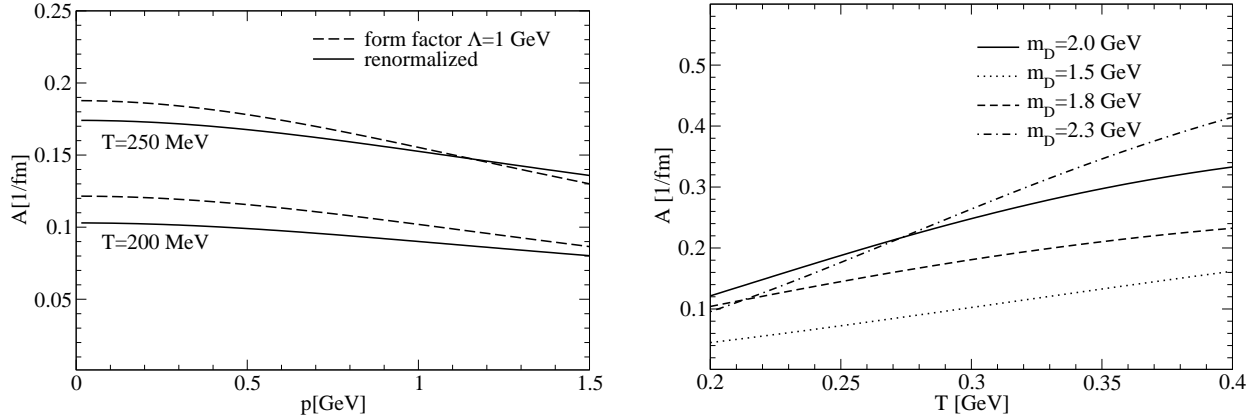

\begin{minipage}{0.48\textwidth}
\includegraphics[width=\textwidth]{reso-model-form-A}
\end{minipage}\hfill
\begin{minipage}{0.48\textwidth}
\includegraphics[width=\textwidth]{mass-compar}
\end{minipage}
\caption{Left panel: Comparison of the friction coefficient for resonances
  with self-energies calculated in the renormalization scheme
  (\ref{rencond}) (solid line) and with form-factor regularization (dashed
  line). The coupling was chosen such that in both the resonance width
  $\Gamma$=0.4~GeV at the resonance mass. Right panel: The friction
  coefficient for different resonance masses as function of the
  temperature, evaluated in the form-factor regularization
  scheme.\label{fig_params}}
\end{figure}

To further illustrate the uncertainties inherent to the resonance
properties, we display in Fig.~\ref{fig_params} the sensitivity of the drag
coefficient with respect to the regularization schemes and the resonance
masses (for form-factor regularization). From the left panel one observes
that the momentum dependence of the $A$ coefficient is somewhat more
pronounced for the form-factor regularization, but the absolute magnitude
in both schemes is very similar. The right panel indicates a more
pronounced sensitivity to the value of the resonance mass. In particular,
it confirms that $D$-meson masses close to threshold lead to a significant
reduction of the drag effect, due to the thermal motion of the light
partons from the heat bath. For the same reason, larger masses imply a
steeper increase of $A$ with temperature (of course, above $T\simeq 2T_c$
the very existence of resonance correlations is questionable).

%%%%%%%%%%%%%%%%%%%%%%%%%%%%%%%%%%%%%%%%%%%%%%%%%%%%%%%%%%%%%%%%%
\subsection{Time Evolution of Momentum Spectra}
\label{sec_evo}
%%%%%%%%%%%%%%%%%%%%%%%%%%%%%%%%%%%%%%%%%%%%%%%%%%%%%%%%%%%%%%%%
\subsubsection{Time Dependence of the Fokker Planck Equation}
%%%%%%%%%%%%%%%%%%%%%%%%%%%%%%%%%%%%%%%%%%%%%%%%%%%%%%%%%%%%%%%%

To obtain a better estimate of the effect of heavy-quark rescattering on
the transverse-momentum ($p_T$) spectra of heavy-flavor hadrons in URHICs
we investigate in this section the time evolution of the Fokker-Planck
equation in a thermally evolving QGP. The latter is modelled by an
expanding fireball under conditions resembling central Au-Au collisions
at RHIC. The main simplifying assumption consists of momentum-independent
drag and diffusion coefficients.
According to the discussion at the end of the previous section, we will
therefore employ $D$-meson self-energies within the renormalization scheme
(\ref{rencond}) (with masses and widths of $m_D$=2~GeV and
$\Gamma$=0.4~GeV), as well as pQCD cross sections (with $\alpha_s$=0.4),
cf. Fig.~\ref{fig_coeff}.  The Fokker-Planck Eq.~(\ref{fp-eq}) then takes
the form
\begin{equation}
\label{fp-const-coeff}
\frac{\partial f}{\partial t} = \gamma(t) \frac{\partial}{\partial \vec{p}}
(\vec{p} f) + D(t) \frac{\partial^2}{\partial \vec{p}^2} f \ ,
\end{equation}
where $\gamma=A(T(t),|\vec p|=0)$ and 
$D=B_0(T(t),|\vec p|=0)=B_1(T(t),|\vec p|=0)$.
The time dependence of the coefficients enters through their dependence 
on temperature (as determined in the previous section), with $T(t)$ 
following from the fireball model outlined below. For the initial 
condition,
 \begin{equation}
\label{f0}
f(t=0,\vec{p}) \equiv f_0(\vec{p}) \ ,
\end{equation}
we will employ $c$-quark spectra from $p$-$p$ collisions as extracted
from the PYTHIA event generator~\cite{pythia01,greco-private}.
The initial-value problem can be conveniently solved employing Green's 
function techniques: if we can find a solution $G(t,\vec{p};\vec{p}_0)$
to Eq.~(\ref{fp-const-coeff}) with the initial condition
\begin{equation}
\label{fp-green}
G(t=0,\vec{p};\vec{p}_0) = \delta^{(3)}(\vec{p}-\vec{p}_0) \ ,
\end{equation}
the full solution with an arbitrary initial condition (\ref{f0}) follows as
\begin{equation}
\label{fp-sol-green}
f(t,\vec{p})=\int \d^3 \vec{p}_0 G(t,\vec{p};\vec{p}_0) f_0(\vec{p}_0) \ .
\end{equation}
To determine the Green's function, we define its Fourier transform,
\begin{equation}
\label{fp-four}
G(t,\vec{p};\vec{p}_0) = \int d^3 \vec{q} \ 
\exp(-\ii\vec{q}\cdot\vec{p})  \ g(t,\vec{q};\vec{p}_0) \ ,
\end{equation}
and insert it into Eq.~(\ref{fp-const-coeff}), leading to the first-order 
differential equation 
\begin{equation}
\label{fp-four-eq}
\frac{\partial g}{\partial t}+\gamma \vec{q} \frac{\partial g}{\partial
  \vec{q}} = - D \vec{q}^2 g  \ .
\end{equation}
With the initial condition for $g$ determined by Eq.~(\ref{fp-green}), 
\begin{equation}
\label{char3}
g(0,\vec{q},\vec{p}_0)=\frac{1}{(2 \pi)^3} 
\exp(-\ii\vec{p}_0\cdot\vec{q}) \ ,
\end{equation}
its solution reads
\begin{equation}
\label{char4}
g(t,\vec{q};\vec{p}_0)=\frac{1}{(2 \pi)^3} 
\exp \left \{\ii \vec{p}_0\cdot\vec{q}
\exp[-\Gamma(t)] \right \} \exp [-\Delta(t) \vec{q}^{~2}] \ ,
\end{equation}
where
\begin{align}
\Gamma(t) &= \int_{0}^{t} \d \tau \gamma(\tau)
\label{gamt}\\
\Delta(t) &= \exp[-2 \Gamma(t)] \int_0^{t} \d \tau D(\tau) 
\exp[2 \Gamma(\tau)] \ .
\label{Dt}
\end{align}
The Fourier transformation (\ref{fp-four}) 
yields the result for the Green's function, 
\begin{equation}
\label{char6}
G(t,\vec{p};\vec{p}_0) = \left [\frac{1}{4 \pi \Delta(t)} \right
  ]^{3/2} \exp \left\{-\frac{(\vec{p}-\vec{p}_0
      \exp[-\Gamma(t)])^2}{4\Delta(t)} \right \} \ , 
\end{equation}
and the time evolution of the distribution function, 
Eq.~(\ref{fp-sol-green}), is finally obtained by numerical integration.

%%%%%%%%%%%%%%%%%%%%%%%%%%%%%%%%%%%%%%%%%%%%%%%%%%%%%%%%%%%%%%%%
\subsubsection{Limiting Cases}
\begin{figure}
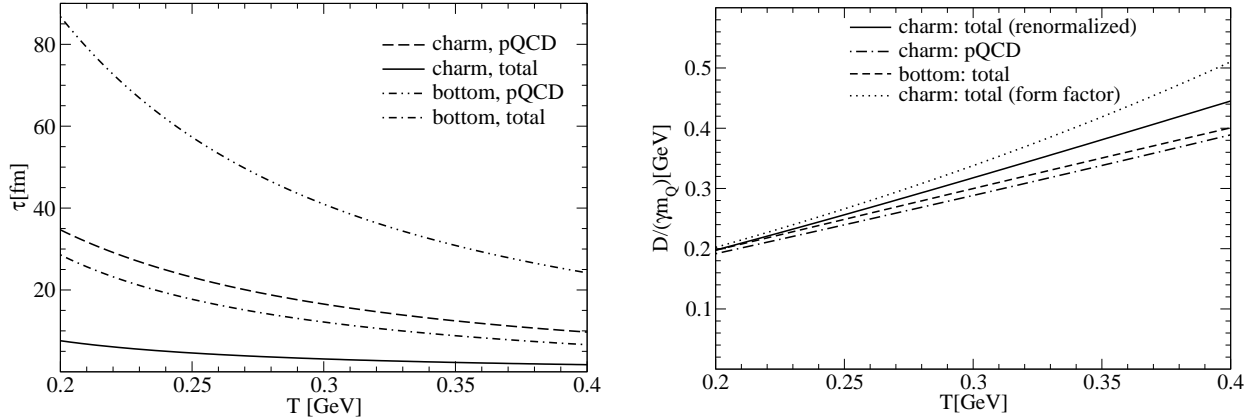

\begin{minipage}{0.48\textwidth}
\includegraphics[width=\textwidth]{tau-charm-bottom}
\end{minipage}\hfill
\begin{minipage}{0.48\textwidth}
\includegraphics[width=\textwidth]{temperature-consist}
\end{minipage}
\caption{Left panel: equilibration time-scale $\tau=1/\gamma$ for charm and
  bottom quarks in the QGP as function of temperature with (solid and 
  dashed line) and without (dash-dotted and dash-double-dotted lines)
  in-medium resonances. Right panel: consistency check of the
  dissipation-fluctuation relation, Eq.~(\ref{diss-fluct}), for
  $c$-quarks with (solid line) and without (dash-dotted line) 
  resonances, as well as for $b$-quarks (with resonances, dashed line),
  in the QGP.} 
\label{fig_limit}
\end{figure}

%%%%%%%%%%%%%%%%%%%%%%%%%%%%%%%%%%%%%%%%%%%%%%%%%%%%%%%%%%%%%%%%
Before we turn to the full solution, let us first illustrate a few 
limiting cases. For time-independent coefficients $\gamma$ and $D$,
Eqs.~(\ref{gamt}) and (\ref{Dt}) simplify to
\begin{equation}
\Gamma(t)=\gamma t \ , \quad 
\Delta(t)=\frac{D}{2 \gamma} [1-\exp(-2 \gamma t)] \ . 
\end{equation}
Inserting these into Eq.~(\ref{char6}) yields
\begin{equation}
\label{fp-sol}
G(t,\vec{p};\vec{p}_0) = 
\left \{ \frac{\gamma}{2 \pi D[1-\exp(-2 \gamma t)]} \right \}^{3/2} 
\exp \left \{-\frac{\gamma}{2D} \frac{[\vec{p}-\vec{p}_0
      \exp(-\gamma t)]^2}{1-\exp(-2 \gamma t)}\right \},
\end{equation}
which was already derived in Ref.~\cite{Svet88}.  This, in particular,
shows that $\gamma$ has the meaning of a drag (or friction) coefficient,
\begin{equation}
\label{fp-drag}
\erw{\vec{p}(t)}=\vec{p}_0 \exp(-\gamma t) \ ,
\end{equation}
characterizing the equilibration time scale $\tau=1/\gamma$. With the
momentum fluctuation given by
\begin{equation}
\label{fp-diff}
\erw{\vec{p}^{\,2}(t)}-\erw{\vec{p}(t)}^2
=\frac{3D}{\gamma} [1-\exp(-2\gamma t)] \ ,
\end{equation}
$D$ is readily identified as a momentum diffusion coefficient. From the
left panel of Fig.~\ref{fig_limit} we see that for temperatures expected 
in central Au-Au collisions at RHIC, the equilibration time scale for
charm quarks is substantially reduced by resonant interactions, to about a
few fm/$c$, comparable to the duration of the (putative) QGP phase. Even 
though a
similar mechanism is operative for bottom quarks, their much larger rest
mass renders the pertinent kinetic equilibration time significantly larger,
around 10~fm/$c$ or more.

For $t \rightarrow \infty$, Eq.~(\ref{fp-sol}) approaches a 
Maxwell-Boltzmann distribution,
\begin{equation}
\label{fp-equil}
\lim_{t \rightarrow \infty} G(t,\vec{p};\vec{p}_0) =
f_{\text{eq}}(\vec{p})=\left (\frac{\gamma}{2 \pi D} \right)^{3/2} \exp 
\left [-\frac{\gamma \vec{p}^{\,2}}{2D} \right ] \ ,
\end{equation}
and thermal equilibrium implies the dissipation-fluctuation theorem,  
\begin{equation}
\label{diss-fluct}
T = \frac{D}{\gamma m_c} \ .
\end{equation}
The such obtained temperature $T$ should, of course, coincide with the one
of the (light-quark and gluon) heat bath entering through the thermal
distribution functions in Eq.~(\ref{erw}). Thus, Eq.~(\ref{diss-fluct})
serves as a consistency check for the determination of the drag and
diffusion coefficients within our model, especially for the assumption on
the dominance of small momentum transfers underlying the Fokker-Planck
Equation (\ref{fp-eq}). From the right panel in Fig.~\ref{fig_limit} we see
that, for charm quarks, the dissipation-fluctuation theorem is well
satisfied (within 3\%) when using forward peaked pQCD cross sections; but
also for the isotropic resonance cross sections within the renormalization
scheme the deviations do not exceed 11\% even at the highest considered
temperatures, while within the form-factor regularization scheme they reach
up to 26\% (the latter value is reduced to $\sim$17\% when an average over
a thermal momentum distribution is performed). The latter is due to the
greater variation of the friction and diffusion coefficients with momentum
which makes the approximation of momentum-independent coefficients,
underlying the derivation of the fluctuation-dissipation relation
(\ref{diss-fluct}), less accurate. For this reason, in the next section we
shall use the renormalization scheme without form factor to investigate the
time evolution of $p_T$ spectra.  For the heavier bottom quarks the
fluctuation-dissipation theorem is satisfied to high accuracy.

%%%%%%%%%%%%%%%%%%%%%%%%%%%%%%%%%%%%%%%%%%%%%%%%%%%%%%%%%%%%%%%%
\subsubsection{Charm-Quark $p_T$-Spectra at RHIC}
%%%%%%%%%%%%%%%%%%%%%%%%%%%%%%%%%%%%%%%%%%%%%%%%%%%%%%%%%%%%%%%%

Let us finally address the time evolution of charm-quark $p_T$-spectra
including the temperature dependence of drag and diffusion coefficients. To
obtain the time evolution of the temperature we use a simple expanding
fireball model~\cite{Rapp01}.  In reminiscence to hydrodynamic
simulations~\cite{kolbrap03} of central Au-Au collisions at RHIC, the
fireball volume is parameterized by
\begin{equation}
\label{fireball}
V_{\text{FB}}(t)=\pi (z_0+v_z t)(r_0+\frac{1}{2} a_{\perp} t^2)^2,
\end{equation}
where $r_0=6.5$~fm and $z_0=0.6$~fm are the initial transverse and
longitudinal size (the latter corresponding to a formation time of
$\tau_0=0.33$~fm/$c$). The longitudinal and transverse expansion
are characterized by $v_z=1.4c$ (covering a thermal width of about
1.8 units in rapidity) and $a_{\perp}=0.055$~$c^2$/fm (yielding
a total fireball lifetime of about 14~fm/$c$ with a thermal freezeout 
temperature of $\sim$110~MeV). 

Assuming isentropic expansion, the temperature at each instant is
calculated from the total entropy of produced particles, $S=s(T)
V_{\text{FB}}(t)\simeq10^4$ (within $\Delta y$=1.8), with the entropy 
density in the QGP given by
\begin{equation}
\label{entdens}
s=\frac{4 \pi^2}{90} T^3 (16+10.5 N_f) \ ,
\end{equation}
($N_f$ is the effective number of quark flavors, taken to be 2.5).

With these parameters, the initial QGP temperature is $T_0\simeq375$~MeV,
decreasing to the critical temperature of $T_c\simeq180$~MeV after about
$3$~fm/$c$, with further evolution in a hadron-QGP mixed phase for another
$3$~fm/$c$. For simplicity, we treat the latter phase as a QGP at constant
$T=T_c\simeq 180$~MeV, but accounting for lower parton densities as
estimated from the temperature dependence of the Fokker-Planck coefficients
(from the lower panel of Fig. \ref{fig_coeff} we estimated $A,B \propto
\varrho^{2/3}$, where $\varrho$ denotes the density of the light partons).

For the initial distribution $f_0(\vec p)$, Eq.~(\ref{f0}), we employ
charm-quark $p_T$-spectra as generated in proton-proton ($p$-$p$)
collisions at 200~GeV by PYTHIA~\cite{pythia01}.  A suitable
parameterization thereof is given by~\cite{greco-private}
\begin{equation}
\label{pTdist}
\frac{\d^2 N_c}{\d p_{T}^2} = C \frac{(p_T+A)^2}{(1+p_T/B)^{\alpha}}
\end{equation}
with $A=0.5$~GeV, $B=6.8$~GeV, $\alpha$=$21$, and $C$=$0.845$ GeV$^{-4}$. 

As detailed above, the time evolution of the $p_T$-distribution is obtained
from Eq.~(\ref{fp-sol-green}) with the Green's function (\ref{fp-sol}),
integrated over $p_z$, which amounts to a two-dimensional solution of the
Fokker-Planck equation.
\begin{figure}
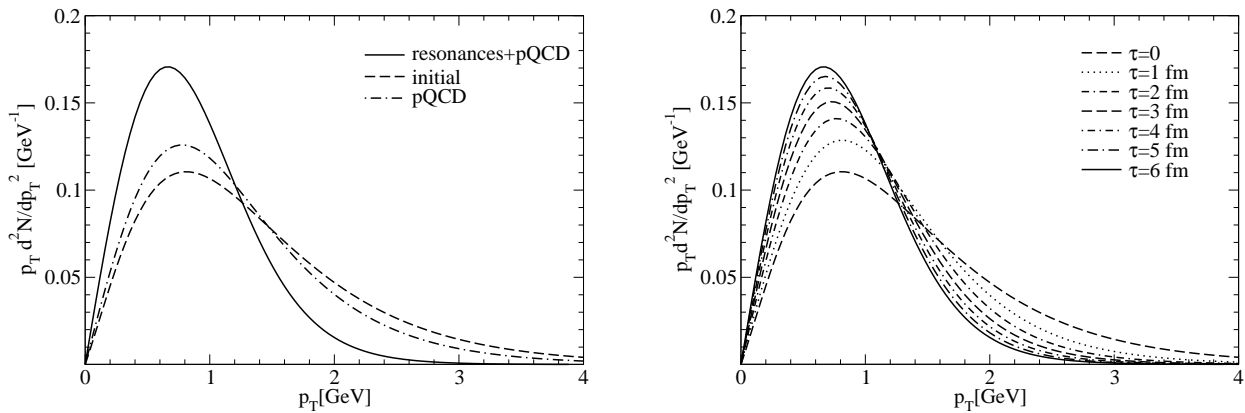

\begin{minipage}{0.47\textwidth}
\includegraphics[width=\textwidth]{time-evol}
\end{minipage}\hfill
\begin{minipage}{0.47\textwidth}
\includegraphics[width=\textwidth]{time-steps}
\end{minipage}
\caption{Left panel: results for the time-evolved $c$-quark $p_T$-spectra
  in the local rest frame with a temperature profile corresponding to QGP
  and mixed phase in central Au-Au collisions at
  $\sqrt{s_{NN}}$=200~GeV; dashed curve: initial spectrum taken from
  $p$-$p$ collisions; dash-dotted curve: final spectrum using pQCD cross
  sections only; solid curve: final spectrum using both pQCD and $D$-meson
  resonance interactions.  Right panel: explicit time evolution in time
  steps of 1~fm/$c$ for the pQCD+resonance interactions.}
\label{fig_tevo}
\end{figure}
Fig.~\ref{fig_tevo} shows the time evolution of $c$-quark $p_T$-spectra in
central Au-Au at RHIC, evaluated in the local (thermal) rest frame of
the expanding matter in QGP and mixed phase (\textit{i.e.}, the additional
boost from the collective transverse expansion is not included), using the
drag and diffusion coefficients computed in Sec. \ref{sec_fp}. From the
left panel one finds that, when allowing for pQCD rescattering only, the
initial spectra from $p$-$p$ collisions are affected rather little.  On the
contrary, when augmenting the interactions with $D$-meson
resonances\footnote{Note that, when adding the matrix elements for pQCD and
  resonance interactions, we did not resum the perturbative contributions
  to, say, $c+\bar q \to c+\bar q$ in the $D$-meson resonance propagator.
  In principle, this would lead to a slight renormalization of the
  resonance mass and width, which, however, is well inside the range of
  uncertainties of the in-medium resonance parameters.}, the $p_T$-spectra
undergo a marked reshaping, essentially a redistribution from high to low
$p_T$.  The final spectra (\textit{i.e.}, at the end of the mixed phase
after 6~fm/$c$), with a peak position at $p_T^{(\text{max})} \simeq
0.66$~GeV, indeed closely resemble a thermal distribution with a
temperature of about $290$~MeV. Even though this indicates that the spectra
are not fully thermalized at $T_c$, the change from an initial average
$\sqrt{\langle p_T^2 \rangle} = 1.66$~GeV (corresponding to a
``temperature" of $\sim$920~MeV) is appreciable.  The fact that most of
rescattering occurs in the early evolution phase (the first 3~fm/$c$ or so,
cf.~right panel of Fig.~\ref{fig_tevo}) should provide favorable conditions
for the build-up of \emph{elliptic} flow.

We recall that our treatment becomes unreliable toward high $p_T$, 
since (i) we have neglected the momentum dependence of the drag and 
diffusion coefficients (\textit{i.e.}, used their values at zero 
momentum), whereas in reality they decrease with $p_T$ 
(cf.~Fig.~\ref{fig_coeff}), and 
(ii) we have not accounted for induced gluon emission which is
expected to be the main mechanism for energy loss of high-$p_T$
partons within pQCD~\cite{djo04,ASW04}. 
Also note that transverse-flow effects, \textit{i.e.}, Lorentz boosts 
of the (partially) thermalized $c$-quarks from the comoving (thermal)
frame into the Lab frame, are not included in Fig.~\ref{fig_tevo}.  
Finally, the (possibly gradual) disappearance of resonance states 
toward high temperatures has not been incorporated (whether this 
happens under RHIC conditions with initial temperatures of 2~$T_c$ 
is not clear at present; \textit{e.g.}, in the lattice calculations
of Refs.~\cite{AH03,KL03} resonance signals are still observed around
these temperatures).

%%%%%%%%%%%%%%%%%%%%%%%%%%%%%%%%%%%%%%%%%%%%
\section{Conclusions and Outlook}
\label{sec_concl}
%%%%%%%%%%%%%%%%%%%%%%%%%%%%%%%%%%%%%%%%%%%%

In the present work we have studied the role of resonant rescattering for
heavy quarks in a Quark-Gluon Plasma at moderate temperatures. Our main
assumption has been the existence of $D$-meson-like resonance states above
$T_c$ which finds support in both effective quark models and recent QCD
lattice calculations. The underlying Lagrangian embodied both chiral and
heavy-quark symmetry, where the latter has been essential to ensure
conserved vector currents (for $D^*$ resonances). Bare masses and coupling
constants of the model have been adjusted to render one-loop resummed
$D$-meson spectral functions reminiscent to more microscopic calculations.

Pertinent cross sections for resonant $c$-quark rescattering on light
antiquarks in the heat bath have been applied within a Fokker-Planck
equation to evaluate kinetic thermalization. Our main finding is that the
introduction of resonances in the QGP leads to a substantial reduction of
the equilibration-time scales (by a factor of $\sim$3) as compared to
estimates based on perturbative interactions.  To a large extent, this
difference originates from the isotropy of the angular distributions for
resonant scattering, as opposed to mostly forward scattering in pQCD, which
importantly enters into the (angular-weighted) transport cross section.
The reduced timescales are very reminiscent to expected QGP lifetimes in
central Au-Au collisions at RHIC.  Consequently, time-evolved $c$-quark
$p_T$-spectra exhibit a marked tendency toward thermalization in the
resonance picture, whereas they are only little affected by pQCD
rescattering alone. This should have important consequences for the
build-up of elliptic flow of charmed hadrons.

Let us briefly discuss further ramifications and directions for future
work. Clearly, the three-momentum dependence of the drag and diffusion
coefficients, the effects of transverse flow and the evaluation of elliptic
flow need to be addressed. Quasiparticle masses for light quarks and gluons
should be introduced to make a closer connection to the QGP equation of
state. The disappearance of the bound states needs to be accounted for,
especially for applications at LHC with larger anticipated initial
temperatures than at RHIC. Indeed, a key question that may eventually be
answered by lattice QCD is whether $D$-meson (and other) resonances in the
QGP exist, and, if so, whether they are located \emph{above} the two-quark
threshold, which renders them accessible for direct ($c$+$\bar
q$$\to$$c$+$\bar q$) scattering processes.  Clearly, if this is not the
case, other processes, such as $c$+$\bar q$$\to$$D$+$g$, need to be
evaluated. The effects on secondary $c\bar{c}$ production should also be
checked~\cite{lev94}.  \textit{E.g.}, according to recent transport
calculations~\cite{Mol04}, upscaling pQCD cross sections by a factor of 3
(to generate a significant elliptic flow) entails an increase in open-charm
pairs by 40-50\% over primordial production in central Au-Au collisions
at RHIC, which is not supported by current PHENIX data~\cite{adler04}.
Resonance cross sections, due to the larger $D$-meson mass in the crossed
channel, may not have this feature (or, at least, to a less extent).
Finally, we recall the important impact that heavy-quark momentum
distributions have on secondary production (``coalescence'') of charmonium
and bottomonium states.  Obviously, the combined theoretical and
experimental study of heavy-flavor probes promises a rich potential for the
understanding of the complex nature of QGP at moderate temperatures.

\section*{Acknowledgments}

One of us (HvH) thanks the Alexander von Humboldt foundation for support
within a Feodor Lynen fellowship. This work was supported in part
by a U.S. National Science Foundation CAREER award under grant 
PHY-0449489.

\end{document}